\documentstyle[epsf,twocolumn,prl,aps]{revtex}

\newcommand{\tj}{$t$-$J$\ }

\newcommand{\etal}{{\it et al}}
\newcommand{\ij}{\langle ij\rangle}
\renewcommand{\S}{{\vec S}}  
\newcommand{\grapprox}{\stackrel{>}{_\sim}}
\newcommand{\lapprox}{\stackrel{<}{_\sim}}

\begin{document}
\draft

 \twocolumn[\hsize\textwidth\columnwidth\hsize\csname @twocolumnfalse\endcsname

\title{Energetics of Domain Walls in the 2D $t$-$J$ model}
\author{ Steven R.\ White$^1$ and D.J.\ Scalapino$^2$}
\address{ 
$^1$Department of Physics and Astronomy,
University of California,
Irvine, CA 92697
}
\address{ 
$^2$Department of Physics,
University of California,
Santa Barbara, CA 93106
}
\date{\today}
\maketitle
\begin{abstract}
\noindent 

Using the density matrix renormalization group,
we calculate the energy of a domain wall in the 2D \tj\ model
as a function of the linear hole density $\rho_\ell$, as well as the
interaction energy between walls, for $J/t=0.35$.
Based on these results, we conclude that the ground state always
has domain walls for dopings $0 < x \lapprox 0.3$.
For $x \lapprox 0.125$, the system has (1,0) domain walls with $\rho_\ell
\sim 0.5$, while for $0.125 \lapprox x \lapprox 0.17$, 
the system has a possibly 
phase-separated mixture of walls with $\rho_\ell \sim 0.5$ and 
$\rho_\ell =1$. For $x \grapprox 0.17$, there are only walls 
with $\rho_\ell =1$.
For $\rho_\ell = 1$, diagonal (1,1) domain walls have very 
nearly the same energy as (1,0) domain walls.

\end{abstract}

 ]

In the last few years experimental evidence for stripe formation
in the cuprates has been mounting\cite{stripeexper}. A proper
theoretical description of domain walls and striped phases in
a doped two-dimensional antiferromagnet has
been extremely difficult to develop, however. Although simple
mean field theories for the \tj or Hubbard models yield domain
walls, it is clear that real domain walls have a much more
subtly correlated ground state. Partially filled 
domain walls have been particularly hard to describe
theoretically.

Recently, we reported numerical results\cite{stripe} showing a
striped phase in a $16\times8$ \tj system at a filling of
$x=0.125$, which were in agreement with neutron scattering
results for La$_{1.6-x}$Nd$_{0.4}$Sr$_x$CuO$_4$, a system in
which a suppression of superconductivity occurs near
$x=0.125$.  Here, domain
walls with a linear filling of 1/2 hole per unit length
separating $\pi$-phase-shifted antiferromagnetic regions were
spaced four lattice spacings apart. The question of what happens
at other fillings was not addressed, and the possibility of
other types of domain walls, such as diagonal walls, was
not considered. 
Experimentally, in the Nd$_{0.4}$ system Tranquada et.
al.\cite{stripeexper}
report coexistence of superconducting and domain order for a
range of dopings away from $x=0.125$. For
$0.05 \lapprox x \lapprox 0.12$, the inverse domain spacing was
found to vary as $2 x$. Beyond $x=0.12$ the inverse spacing
remained relatively constant, increasing slightly as $x$
approached 0.2.

Here we address the question of the stability of domain walls at
low to moderate filling, by calculating the energy of a domain wall
in the 2D \tj model, as a function of filling, using
density matrix renormalization group (DMRG)\cite{dmrg}
techniques.
We use systems with boundary conditions (BCs) carefully
chosen not to frustrate the domain walls. We also estimate
the repulsive interaction between domain walls, allowing us
to study the domain-wall filling and spacing of a striped phase
as a function of doping $x$.
The DMRG results for energies use extrapolation to extract
the limit of zero truncation error\cite{hubthreechain}, with up
to 1400 states per block kept. More details of the numerical
techniques can be found in \cite{stripe}.

The \tj\ Hamiltonian in the subspace of no doubly occupied sites
is given by
\begin{equation}
H = - t \sum_{\langle ij \rangle s}
      ( c_{is}^{\dagger}c_{js}
                + {\rm h.c.}) + 
J \sum_{\langle ij \rangle}
      ( {\bf S}_{i} \! \cdot \! {\bf S}_{j} -
         \frac{n_i n_j}{4} ) .
\label{tj-ham}
\end{equation}
Here $\ij$ are near-neighbor sites, $s$ is a spin index, $\S_i$
and $c^\dagger_{i,s}$ are electron spin and creation operators,
and $n_i= c^\dagger_{i\uparrow}c_{i\uparrow} +
c^\dagger_{i\downarrow}c_{i\downarrow}$.
The near-neighbor hopping and exchange interactions are $t$ and $J$.  
We measure energies in units of $t$. 
We consider only $J/t=0.35$ here.

First, we consider the energetics of a single domain wall.
Imagine a single long domain wall in the form of a closed loop, with
a fixed number of holes. The loop would be a large rectangle, if
domain walls prefer to be oriented in the (1,0) or (0,1)
directions. If we assume that the domain wall is stable against
evaporation into holes or pairs, then the loop will adjust its
size in order to minimize its energy. The linear hole density
$\rho_\ell$ of the domain wall will have an optimal value
$\bar\rho_\ell$.  We expect that at very low doing, any domain
wall will have doping $\bar\rho_\ell$.  At higher doping,
repulsion between domain walls could lead to increased values of
$\rho_\ell$.

Let $e(\rho_\ell)$ be the energy per hole of a domain wall with density
$\rho_\ell$. Then $\bar\rho_\ell$ minimizes $e(\rho_\ell)$. In order to
measure $e(\rho_\ell)$, some care is needed. The important point is that
a loop can shrink or contract without inducing frustration in the
antiferromagnetic region inside or out. Therefore, in order to reduce finite
size effects, we define $e(\rho_\ell)$ using two systems, each without
frustration. The first system, with energy $E_{\rm dw}$, has a domain
wall and BCs favoring the $\pi$-phase shift of the
antiferromagnetism induced by the domain wall, while the second,
with energy $E_0$, has no holes at all and BCs favoring no
$\pi$-phase shift, but which are otherwise identical. 
Defining $N_{\rm dw}$ as the number of holes
in the domain wall, we have
\begin{equation}
e(\rho_\ell) = \frac{E_{\rm dw}-E_0}{N_{\rm dw}}
\end{equation}
In practice, we use open BCs, with staggered fields
on the edges on either side of the domain wall 
to induce the desired antiferromagnetic order.

\begin{figure}[ht]
\epsfxsize=2.8 in\centerline{\epsffile{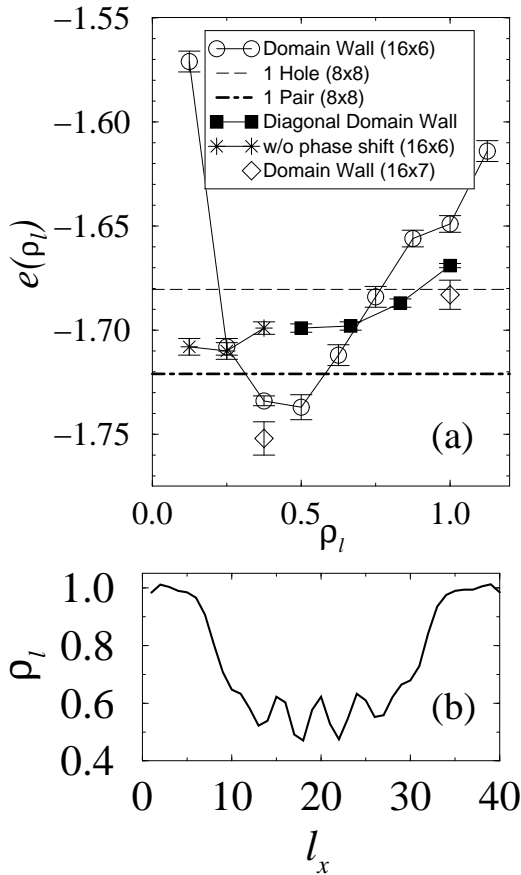}}
\caption{(a) Energy per hole of various hole configurations, as
discussed in the text. (b) Linear density profile for a
$40\times6$ wall, showing phase separation.
}
\end{figure}

In Fig. 1(a) we show $e(\rho_\ell)$ as a function of $\rho_\ell$,
measured on a $16\times6$ system with open BCs, 
with the domain wall parallel to the $x$-axis,
and with staggered magnetic fields
of magnitude $0.1$ applied to the top and bottom rows of sites. 
We see a minimum at $\bar \rho_\ell \sim 0.5$. On the same
plot, we show by horizontal lines the energy per hole of one
hole and of two holes placed in an open $8\times8$ system, 
with a staggered field of magnitude $0.1$ on all four sides.
The fact that the energy per hole of two holes is lower than that 
of a single hole indicates that two holes pair-bind, in
agreement with exact diagonalization studies\cite{pairbind}. 
However, the
energy of a domain wall at $\bar \rho_\ell$ is even lower,
indicating that domain walls are stable at arbitrarily low
doping. To judge the effects of the finite width of the system,
results are shown for a $16\times7$ system at two values of
$\rho_\ell$.  On these larger systems, the energy of the domain wall is
lower\cite{finitesize}.

The $\rho_\ell=1$ domain wall repels additional holes, leading
to the rapid increase in $e(\rho_\ell)$ for $\rho_\ell>1$.
For $\rho_\ell \lapprox 0.3$, the holes are too far apart to
induce the $\pi$ phase shift of a domain wall. Since our
BCs for the doped system require this phase 
shift, for $\rho_\ell \lapprox 0.3$ the energy per hole is quite high.
In this case BCs without the $\pi$ phase shift
give a lower energy, as shown by the stars. However,
here we find that the holes bind into isolated 
pairs rather than a stripe (not shown). 
The two-hole energy without the $\pi$ phase
shift is higher than the two-hole $8\times8$ line because of larger
finite-size effects on the $16\times6$ system.

The concave nature of the $16\times6$ domain wall energy 
for $0.5 < \rho_\ell<1$ 
suggests that in this region, a long domain wall will phase
separate into a region with $\rho_\ell=1$ and a region with
$\rho_\ell=0.5$. We have directly observed this phase separation
in a long $40\times6$ domain wall. 
Fig. 1(b) shows the density profile along the wall for
$\rho_\ell=0.75$. Here the holes have separated into regions
with $\rho_\ell=1$ and $\rho_\ell \sim 0.5 - 0.6$.

In Fig. 1(a), we also show
results for the energy per hole of a diagonal domain wall, 
using a tilted $12\times7$ system, which includes seven adjacent
(1,1) lines of sites. Similar staggered fields for 
doped and undoped systems were applied as for the (1,0) walls
shown in Fig. 1(a). Here the linear hole density is defined as the
number of holes per $\sqrt{2}$ lattice spacings, so that
$\rho_\ell=1$ corresponds to a filled diagonal wall. The energy
near $\rho_\ell=1$ is slightly less than the energy of the (1,0) wall;
however, the width of the diagonal system is slightly greater.
The result shown for a (1,0) domain wall on a $16\times7$
system shows that 
the energies are actually nearly degenerate at $\rho_\ell=1$. 
At lower values of $\rho_\ell$,
(1,0) walls are lower in energy.

\begin{figure}[ht]
\epsfxsize=2.5 in\centerline{\epsffile{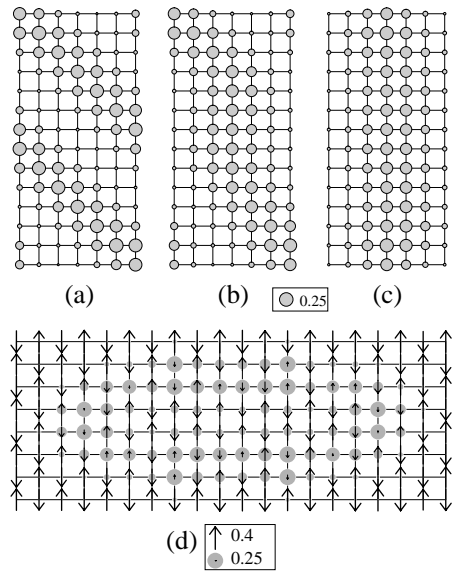}}
\caption{(a)-(c) Hole density, showing domain walls in a 
$14\times7$ system.  The
figures are rotated by $90^\circ$ for compactness. (d) Hole and
spin densities showing a looped
wall on a $20\times8$ system with 16 holes.
The diameter of the gray holes and the length of the arrows are 
proportional to $1-\langle n_i\rangle$ and
$\langle S^z_i \rangle$, respectively,
according to the scales shown.
}
\end{figure}

In Fig. 2(a)-(c) we show a system which allows either a (1,0) or (1,1)
filled domain wall: a $14\times7$ system with cylindrical BCs
(periodic in $y$, open in $x$), doped with 14 holes. Here a
single domain wall extends the length of the system. Staggered
edge fields on the left and right edges and a small
local potential on one site on each edge were used
to pin the ends of the domain wall at specified sites.  
The initial DMRG buildup of the
lattice was arranged to initially force an approximate domain wall of the
specified form to appear; however, later sweeps allowed the
system to relax, although the ends of the walls remained pinned.
Within error bars, all three domain wall configurations shown in 
Fig. 2(a-c) have identical energies. In Fig. 2(a), a (1,1)
domain wall wraps around the system. In Fig. 2(b), a wall with
both (1,0) and (1,1) parts is present. Notice that
the wall resists being situated at an intermediate angle.
In Fig. 2(c), a (1,0) domain wall is present. 
The degeneracy of these states indicates
that filled domain walls have the same energy, whether they
run in the (1,0) or (1,1) directions. This suggests that
filled domain walls might readily fluctuate or 
form static disordered configurations.

Note that at low doping, in a strictly 2D system, one would expect
infinite, straight (1,0) domain walls with filling given by the
optimal filling $\bar \rho_\ell \approx 0.5$. 
However, in a system with weak coupling
to other planes, in order to maintain long range
antiferromagnetic order, it is more likely that domain walls would
form closed loops, so that most of each plane would be in the
dominant antiferromagnetic domain. 
In Fig. 2(d) we show a loop on a $20\times8$
system. In this system, staggered fields without any phase
shifts were applied to all four sides, preventing a domain wall
from ending on a side. Under these circumstances, a loop forms.
In a system of weakly coupled planes, the size of a typical loop
would be set to balance repulsion between opposite sides of
the loop and the exchange cost from the coupling to adjacent
planes. (If the exchange coupling between planes were
$J' \sim 10^{-5} J$, and assuming the repulsion is given by Eq.
(3), the loop size would be about 15-20 lattice spacings.)
At higher dopings of just a few percent, the interactions
between walls would favor the more closely packed striped phases.

In order to understand this interaction between domain walls,
we have studied an $L\times6$ system with cylindrical BCs.
In this system, transverse domain walls with four holes wrap 
around the system, and are stable at low to moderate doping.
These walls have $\rho_\ell=2/3$.
We studied systems with eight or twelve holes, forming two or
three domain walls,
and studied various lengths $L$ to determine the interaction
between the walls. The energy per hole as a function of the
domain wall spacing $d$ is shown in Fig. 3(a). The walls repel, rather
strongly at short distances. The solid curve in Fig. 3(a) is a
simple exponential fit
\begin{equation}
e(d) \equiv  e(\infty) + V(d) = e(\infty) + A e^{-d/w} 
\label{repulsion}
\end{equation}
with $e(\infty) = -1.79$, $A = 0.87$ and $w=1.8$.
The source of the repulsion
appears to be the finite width of the walls: the hole density
distribution spreads out over several lattice spacings. In 
the insert to Fig. 3(a), we show the hole density per site as a function of
$\ell_x$ for a $13\times6$ system with a single domain wall in
the center. This gives the density profile of a wall. An 
isolated wall, far from boundaries in the $L\times6$ system with
$\rho_\ell=2/3$,
is site-centered. However, walls near boundaries can be 
more bond-centered; there appears to be little energy
difference between walls which are site centered, bond centered, or in
between.  Notice the substantial width of the wall; only
$30\%$ of the hole density is on the center leg. The effective
mass of the wall as a whole seems to be very high, so that it
is effectively pinned by truncation errors in DMRG. In other
words, we believe very little of the apparent width shown is due to
uniform motion of the entire wall\cite{casimir}. 

\begin{figure}[ht]
\epsfxsize=2.5 in\centerline{\epsffile{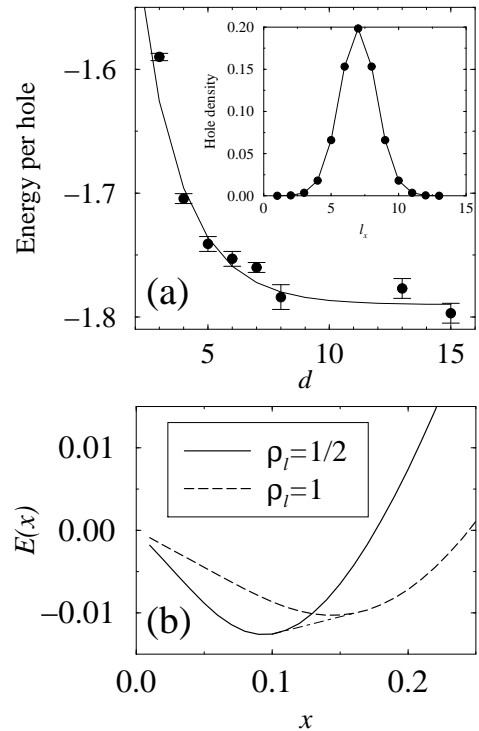}}
\caption{(a) Energy per hole of (0,1) domain walls on a 
$16\times6$ system, as a function of domain wall separation.
The inset shows the transverse hole density in one of the walls.
(b) The energy of an array of domain walls as a function of 
doping $x$. 
}
\end{figure}
 
Using the results of Fig. 1 for the energy per hole in a domain
wall and Eq.(3) for the repulsion per hole between domain
walls, we consider the relationship between the domain wall
spacing $d$ and the doping $x$. 
The energy per site of an array of domain walls is given by
\begin{equation}
E(\rho_\ell,d) \equiv  x [e(\rho_\ell) + V(d)] .
\end{equation}
For $\rho_\ell=1/2$, $d=1/(2x)$, while for $\rho_\ell=1$,
$d=1/x$.  
For fixed $\rho_\ell$, we define 
\begin{equation}
E(x) \equiv  E(\rho_\ell,d(x)) - x [2e(1) - e(0.5)] .
\end{equation}
The latter term in Eq. (5) is like a shift in the chemical
potential, which allows the curvature in the energy to be seen
more easily.
In Fig. 3(b), we plot $E(x)$ for
$\rho_\ell=1/2$ and $\rho_\ell=1$ arrays of walls.
Clearly, at low values of $x$, all the
walls have $\rho_\ell=0.5$ and therefore $d^{-1}=2x$.
For large values of $x$, all the walls have $\rho_\ell=1$ and
therefore $d^{-1}=x$. At intermediate values of $x$,
$0.11 < x < 0.16$, using a
Maxwell construction one finds a mixture of walls with two
different spacings $d_{1/2}$ and $d_1$. However, the precise values
of $d_{1/2}$ and $d_1$ are sensitive to $V(d)$, which was
determined only roughly using walls on an $L\times6$ system with
$\rho_\ell=2/3$. 

\begin{figure}[ht]
\epsfxsize=2.6 in\centerline{\epsffile{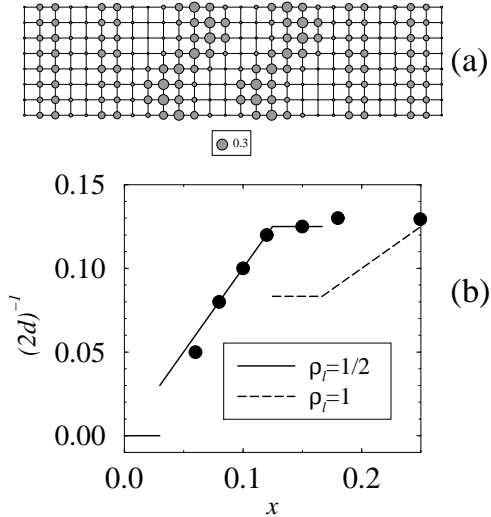}}
\caption{(a) Domain walls with $\rho_\ell=1$ and $\rho_\ell=0.5$ 
on a $28\times8$ system, with 32 holes and cylindrical BCs.
(b) Shift in the magnetic structure factor peak from $(\pi,\pi)$ as
a function of $x$. The solid circles are experimental
results[6] for La$_{2-x}$Sr$_x$CuO$_4$.
}
\end{figure}
 
To determine $d_{1/2}$ and $d_1$ more
precisely, we have simulated a $28\times8$ system, with
cylindrical BCs, with four (0,1)
$\rho_\ell=0.5$ walls and two (0,1) $\rho_\ell=1.0$ walls, 
and $x=0.14$. The results are shown in Fig.  4(a).  
Here, the wall
spacings naturally adjust to $d_{1/2}$ and $d_1$. We find
$d_{1/2} = 4$ and $d_1 = 6$, implying that domain wall phase
separation occurs for $0.125 < x < 0.17$. 
(Note that in this
case, the system cannot continuously adjust $\rho_\ell$ away 
from $1/2$ or $1$, and it may be that interactions between
walls would shift the $\rho_\ell=0.5$ wall to a somewhat
different filling.)
The resulting inverse domain wall spacing is plotted in Fig.
4(b). Also shown are experimental results\cite{yamada} 
for the shift in the magnetic peak in La$_{2-x}$Sr$_x$CuO$_4$.
Here we assume that for $x < 0.03$ the system has loops rather
than stripes so that the magnetic peak would remain at
$(\pi,\pi)$. For $x > 0.125$, walls with $\rho_\ell=1$ and
spacing $d_1$ give rise to an additional peak, which has not
been observed thus far. However, as is obvious from Fig. 4(a),
the $\rho_\ell=1$ walls seem to be much more subject to disorder
and fluctuations, which would significantly weaken and broaden
this peak. Observation of this second peak would lend strong
support to our results, as would a nonmonotonic shift in the
peak near $x \grapprox 0.2$.

In our DMRG calculations, large scale fluctuations
are difficult to observe, so that a slowly fluctuating
striped phase would tend to appear static to us. While this is
certainly a disadvantage of DMRG, it makes it easier to rule out
uniform phases---phases in which there are no signs of domain walls. 
In numerous calculations with various BCs, and
$x < 0.3$, we have not seen a uniform phase\cite{uniform}. 
We have observed apparently disordered walls, particularly with
$x=0.15-0.20$, but it is difficult to determine in these cases whether
an ordered phase is being frustrated by BCs.
Note that even if the $\rho_\ell=1$ walls are fluctuating or
disordered, this probably does not strongly reduce the energy per site 
relative to the slightly disordered walls shown in Fig. 4(a), so
that the phase separation into regions with 
$\rho_\ell=0.5$ walls would be unaffected.

In summary, we find that the (1,0) domain walls formed in
the doped \tj model have a linear filling $\rho_\ell \approx 0.5$ and an
inverse spacing $d^{-1}=2x$ for doping $x\lapprox 0.125$.
For $x \grapprox 0.17$, the domain walls have $\rho_\ell=1$
and $d^{-1}=x$. It is tempting to identify the underdoped regime
with the $\rho_\ell=0.5$ walls and the overdoped regime with 
$\rho_\ell=1$. We also find that $\rho_\ell=1$
(1,0) walls are nearly degenerate with diagonal
(1,1) domain walls, suggesting that these walls may have large
fluctuations.


We acknowledge support from the NSF under 
Grant No. DMR-9509945 (SRW), 
PHY-9407194 (DJS), and DMR-9527304 (DJS).


%
\newpage


\begin{references}
\bibitem{stripeexper} J.M.~Tranquada \etal, Nature {\bf 375}, 561 (1995);
\prb {\bf 54}, 7489 (1996); \prl {\bf 78}, 338 (1997).

\bibitem{stripe}S.R.~White and D.J.~Scalapino, \prl {\bf 80}, 1272 (1998).

\bibitem{dmrg} S.R. White, \prl {\bf 69}, 2863 (1992),
\prb {\bf 48}, 10345 (1993).

\bibitem{hubthreechain} J. Bonca, J.E. Gubernatis, M. Guerrero, Eric 
Jeckelmann, and Steven R. White, condmat 9712018.

\bibitem{pairbind} E. Dagotto, Rev. Mod. Phys.
{\bf 66}, 763 (1994).

\bibitem{finitesize} Given that the finite size corrections
are comparable to the differences in energy, we cannot
conclusively determine that the domain wall phase is stable at
very low doping.  However, as can be seen from the curve in Fig.
1(a) labeled ``w/o phase shift", the energy of the pair phase
rises rapidly with doping.

\bibitem{casimir} For large $d$, the interaction between walls
may become attractive because of a Casimir-like effect.
See L. P. Pryadko, S. Kivelson, and D. W. Hone, cond-mat/9711129.

\bibitem{yamada} K. Yamada, C.H. Lee, Y. Endoh, G. Shirane, R.J.
Birgeneau, M.A. Kastner, Physica C {\bf 282-287}, 85 (1997).

\bibitem{uniform} At $x = 0.5$, we see a quite uniform phase on 
$L\times6$ systems.


\end{references}
\end{document}